\documentclass[twocolumn,prl,showpacs,color,superscriptaddress,epsfig,floatfix,amsmath,amssymb,amsfonts]{revtex4}

\usepackage{subfig}
\usepackage{graphicx}
\usepackage{multirow,color}

\begin{document}

\title{The frustrated Heisenberg antiferromagnet on the honeycomb lattice:
    \\A candidate for deconfined quantum criticality}

\author{D. J. J. Farnell}

\affiliation{Division of Mathematics, University of Glamorgan, Pontypridd CF37 1DL, Wales, UK}

\author{R. F. Bishop}

\affiliation{School of Physics and Astronomy, The University of Manchester, Manchester M13 9PL, UK}

\author{P. H. Y. Li}

\affiliation{School of Physics and Astronomy, The University of Manchester, Manchester M13 9PL, UK}

\author{J. Richter}

\affiliation{Institut f\"ur Theoretische Physik, Otto-von-Guericke Universit\"at Magdeburg, P.O.B. 4120, 39016 Magdeburg, Germany}

\author{C.~E.~Campbell}
\affiliation{School of Physics and Astronomy, University of Minnesota, Minneapolis, Minnesota 55455, USA}

\date{\today}

\begin{abstract}
  We study the ground-state (gs) phase diagram of the frustrated
  spin-$\frac{1}{2}$ $J_{1}$--$J_{2}$--$J_{3}$ antiferromagnet with
  $J_{2}=J_{3}=\kappa J_1$ on the honeycomb lattice, using
  coupled-cluster theory and exact diagonalization methods.  We
  present results for the gs energy, magnetic order parameter,
  spin-spin correlation function, and plaquette valence-bond crystal
  (PVBC) susceptibility.  We find a N\'{e}el antiferromagnetic (AFM)
  phase for $\kappa < \kappa_{c_{1}} \approx 0.47$, a collinear
  striped AFM phase for $\kappa > \kappa_{c_{2}} \approx 0.60$, and a
  paramagnetic PVBC phase for $\kappa_{c_{1}} \lesssim \kappa \lesssim
  \kappa_{c_{2}}$.  The transition at $\kappa_{c_{2}}$ appears to be
  of first-order type, while that at $\kappa_{c_{1}}$ is
  continuous. Since the N\'eel and PVBC phases break different
  symmetries our results favor the deconfinement scenario for the
  transition at $\kappa_{c_{1}}$.
\end{abstract}


\maketitle

Frustrated quantum Heisenberg magnets are paradigms of those systems
that may be used to study quantum phase transitions (QPTs) between
quasiclassical magnetically ordered phases and magnetically disordered
quantum phases.  In particular, if the quasiclassical phase and the
quantum phase spontaneously break different symmetries, the
Landau-Ginzburg-Wilson scenario of continuous phase transitions does not hold
and the concept of deconfined quantum criticality \cite{Sent:2004}
becomes relevant. A canonical model for studying such QPTs is the
spin-$\frac{1}{2}$ Heisenberg antiferromagnet (HAFM) with
antiferromagnetic (AFM) nearest-neighbor (nn) $J_1$ coupling and
frustrating AFM next-nearest-neighbor (nnn) $J_2$ coupling (namely the
$J_1$--$J_2$ model) on the square lattice (see, e.g.,
Refs. \cite{schulz,zhito96,Sir:2006,darradi08,reuther,Richter:2010_ED}).
It is now widely acknowledged that this model undergoes a transition
from a quasiclassical N\'{e}el state at $J_2/J_1 \lesssim 0.4$ to a
quantum paramagnetic (QP) phase, which is not magnetically ordered,
for $0.4 \lesssim J_2/J_1\lesssim 0.6 $, and thence to a collinear
striped phase for $J_2/J_1 \gtrsim 0.6$.  The synthesis of layered
magnetic materials that might be described by the square-lattice
$J_1$--$J_2$ model~\cite{melzi00,rosner02} has also encouraged further
theoretical studies of this model.  Although the square-lattice
spin-$\frac{1}{2}$ $J_{1}$--$J_{2}$ HAFM has been intensively studied
for over 20 years, no consensus yet exists on such fundamental issues
as the nature of the QP phase and of the type of transition into
it. In particular, there is still controversy over whether the scenario
of deconfined criticality holds for this model (see, e.g.,
Refs.~\cite{Sir:2006,darradi08}).

Related magnetic systems  may shed light onto this controversy 
and one strongly related model that has received much attention recently
has been the HAFM on the honeycomb lattice.
One reason for this interest is that a spin-liquid phase 
has been found for the exactly solvable Kitaev model~\cite{kitaev}, in
which the spin-$\frac{1}{2}$ particles reside on a honeycomb lattice.
The honeycomb lattice is also relevant to the very active 
research field of graphene, where Hubbard-like
models on this lattice may be appropriate to describe the
relevant physics~\cite{graphene}. Interestingly, Meng {\it et
 al}. \cite{meng} have very recently found clear evidence that 
the quantum fluctuations are strong enough to
establish an insulating spin-liquid phase between the non-magnetic
metallic phase and the AFM Mott insulator for
the Hubbard model on the honeycomb lattice at moderate values of the
Coulomb repulsion $U$. At very large values of $U$, the latter phase 
corresponds to the HAFM on the bipartite honeycomb lattice that 
contains a ground-state showing N\'{e}el
long-range order (LRO). However, higher order terms in the $t/U$ expansion
of the Hubbard model may lead to frustrating exchange couplings in the
corresponding spin-lattice model  where the HAFM with nn 
exchange is the leading term in the large-$U$ expansion. 
This unexpected result \cite{meng} has 
also stimulated interest in the frustrated HAFM on the honeycomb lattice (see e.g.
Refs.~\cite{Cabra,honey5,lauchli2011,honey7}).   
Indeed, the available literature suggests a frustration-induced QP phase 
for the frustrated spin-$\frac{1}{2}$ HAFM on the honeycomb lattice
\cite{honey1,honey2,honey3,Cabra,honey5,lauchli2011,honey7}. These theoretical 
findings are consistent with the recent experimental observations that the
spin-$\frac{3}{2}$ honeycomb-lattice HAFM Bi$_3$Mn$_4$O$_{12}$(NO$_3$)
shows a spin-liquid-like behavior at temperatures much lower
than the Curie-Weiss temperature \cite{exp}.

One approach to shed additional light on the QP regions of such
$J_{1}$--$J_{2}$ models has been to extend its parameter space by the
inclusion of next-next-nearest-neighbor (nnnn) couplings $J_{3}$ as
well. This yields the $J_{1}$--$J_{2}$--$J_{3}$ model (see, e.g.,
Ref.~\cite{Reuther:2011_J1J2J3mod} and references cited therein),
\begin{equation}
H = J_{1}\sum_{\langle i,j \rangle} \mathbf{s}_{i}\cdot\mathbf{s}_{j} + J_{2}\sum_{\langle\langle i,k \rangle\rangle} \mathbf{s}_{i}\cdot\mathbf{s}_{k} + J_{3}\sum_{\langle\langle\langle i,l \rangle\rangle\rangle} \mathbf{s}_{i}\cdot\mathbf{s}_{k}\,,
\label{eq1}
\end{equation}
where $i$ runs over all lattice sites on the lattice, and $j$ runs over all nn sites, 
$k$ over all nnn sites, and $l$ over all nnnn sites to $i$, respectively, counting each bond once and once only. Each site $i$ of the lattice 
carries a spin-$\frac{1}{2}$ particle with spin operator ${\bf s}_{i}$.

\begin{figure}[t]
\begin{center}
\mbox{
\subfloat{\includegraphics[width=6cm,height=6cm,angle=270]{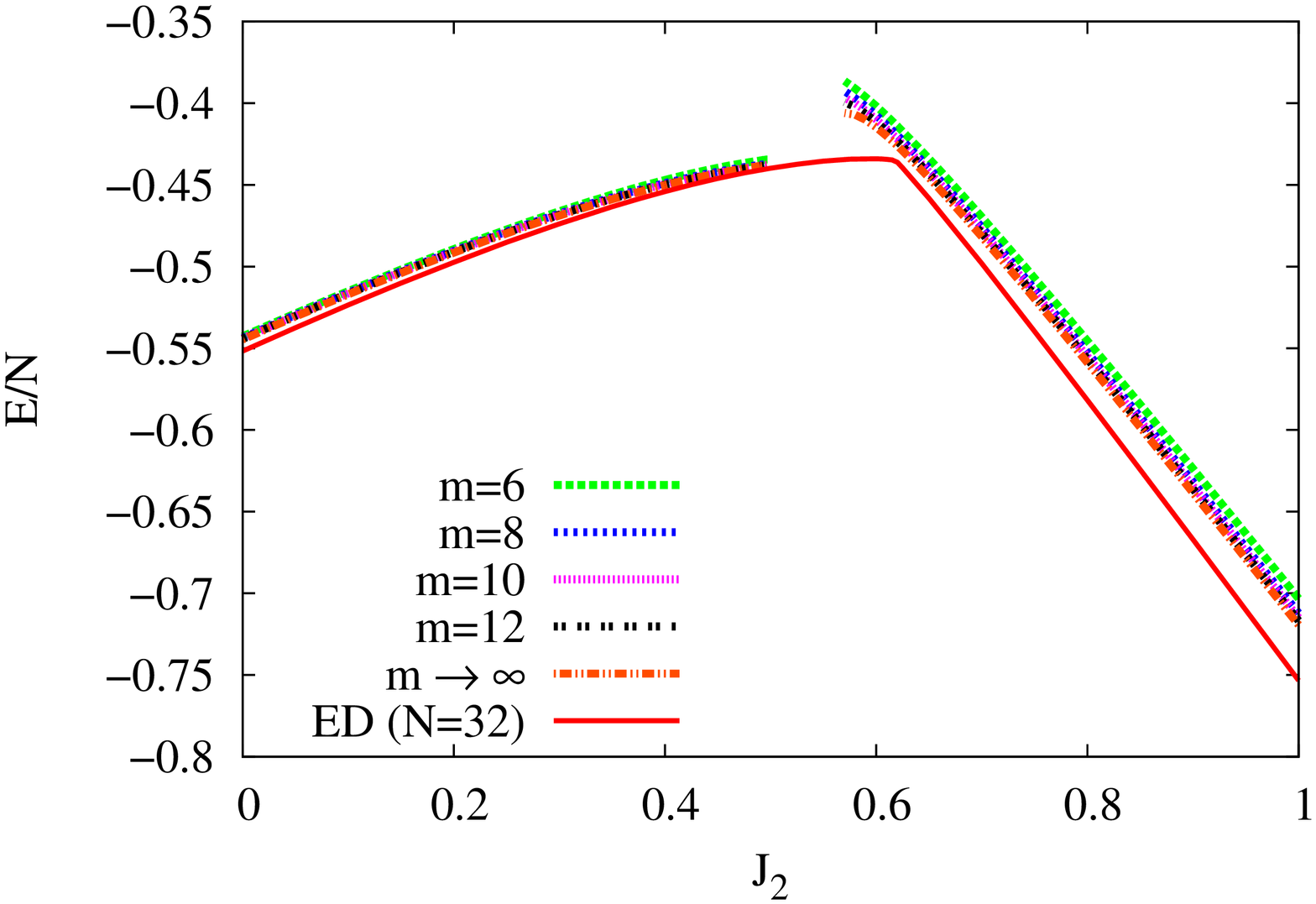}}
\hspace{-1.5cm}
\raisebox{-7.5cm}{
\subfloat{\includegraphics[width=6cm,height=8cm]{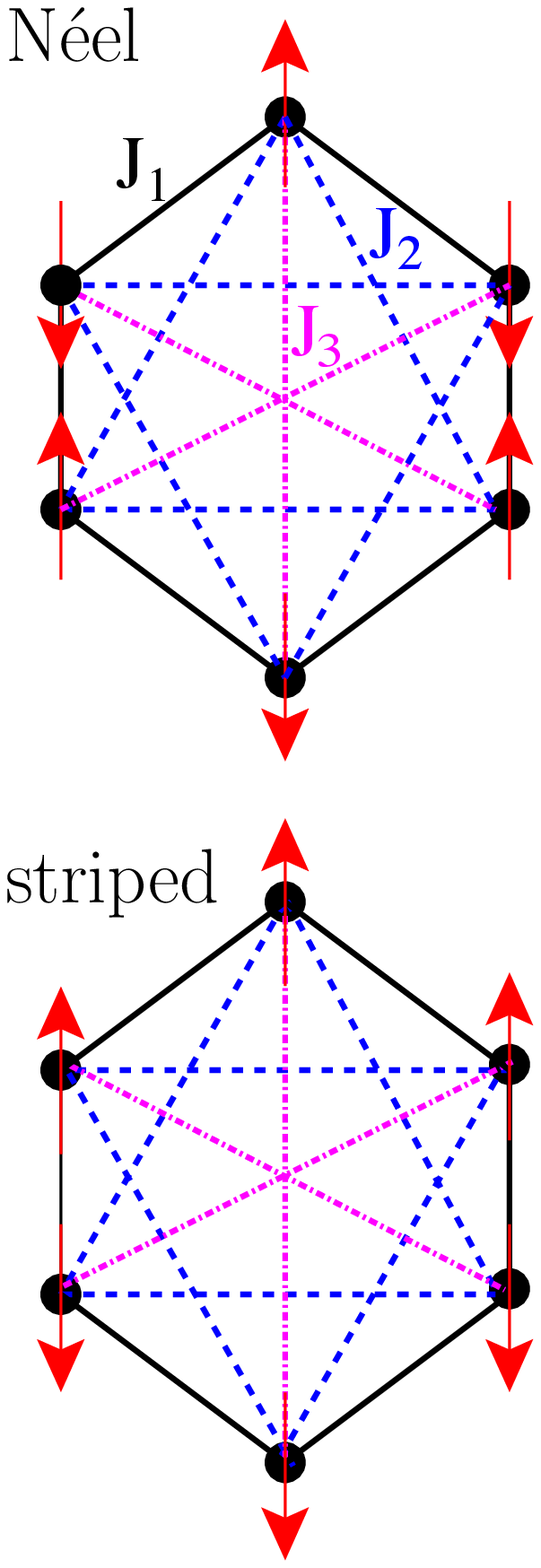}}
}
}
\vskip-2cm
\caption{(Color online) Left: Results for the gs energy, $E/N$ ($J_{1}
  \equiv 1$ and $J_{3}=J_{2}$) via ED ($N=32$), CCM LSUB$m$
  ($m=6,8,10,12$) and extrapolated $m \rightarrow \infty$ (see text).
  Right: The $J_{1}$-$J_{2}$-$J_{3}$ honeycomb model;
  for the N\'{e}el and the collinear striped
  model states.  The arrows represent spins located on specific
  lattice sites [symbolized by \textbullet].}

\label{E}
\end{center}
\end{figure}

Here we wish to study the
spin-$\frac{1}{2}$ $J_{1}$--$J_{2}$--$J_{3}$ HAFM model 
of Eq.~(\ref{eq1}) on the
honeycomb
lattice~\cite{honey1,honey2,honey3,Cabra,honey5,lauchli2011,honey7}. The
lattice and the exchange bonds are illustrated in Fig.~\ref{E} (right
panel),
where the quasiclassical N\'{e}el and collinear striped phases are
also shown.
Classically, this system exhibits the two collinear phases illustrated
in Fig.~\ref{E} (right panel) as well as a spiral phase. These phases
meet in a triple point at $J_2=J_3=J_1/2$, (see, e.g.,
Refs.~\cite{Cabra,honey2}).  Henceforth we set $J_{1} \equiv 1$ and we
further restrict ourselves to the case $J_{2} = J_{3} \equiv \kappa
J_{1}$.  The motivation to focus on $J_{2} = J_{3}$ also comes
from the classical case, where the honeycomb-lattice model and the
square-lattice $J_{1}$--$J_{2}$ HAFM model have similar ground-state
properties, namely: (i) there is a direct transition between the
N\'{e}el and the collinear striped states at $\kappa =0.5$; and,
(ii) the classical ground state is highly degenerate at the transition
point.  Hence, the two systems are very similar classically, and
so the honeycomb system is a likely candidate to exhibit a QP
phase and possibly also a deconfined critical point between this
QP phase and the quasiclassical phases. Moreover, we note again that
non-zero $J_3$ bonds are likely from higher-order terms in the $t/U$
expansion of the Hubbard model at moderate $U$ values \cite{meng}.

The treatment of highly frustrated quantum magnets is notoriously
challenging, since only a relatively few accurate methods exist. 
New promising approaches such as projected entangled pair states \cite{peps} 
are not yet accurate enough.  The finite-size extrapolation of
numerical exact data for finite-lattice systems, used successfully for the
square-lattice model \cite{schulz,Richter:2010_ED}, 
is also somewhat less efficient for the
honeycomb lattice,
since (i) the unit cell contains two sites, (ii) 
nnnn couplings $J_3$ are included,  and (iii) there are only a few finite
lattices with full point group symmetry \cite{honey2}.   
Naturally, quantum Monte Carlo methods are severely
limited in the presence of frustration due to the infamous ``sign problem.''
The coupled-cluster method (CCM) when evaluated to
high orders of approximation provides a reasonably 
accurate approach to determine the position of QPT points
\cite{rachid05,schmalfuss,Bi:2008_PRB,darradi08,richter10,Reuther:2011_J1J2J3mod},
as well as providing evidence on the nature of such QP 
phases \cite{darradi08}.
Here we use this method together with
complementary results provided by the exact diagonalization (ED) technique
for a finite lattice of $N=32$ sites.

The CCM is an intrinsically size-extensive method that 
provides results in the limit $N \rightarrow \infty$
from the outset. The CCM requires us to input a model (or reference)
state~\cite{ccm2,ccm3,j1j2_square_ccm1},
with respect to which
the quantum correlations are expressed. Here we use 
the N\'{e}el and striped model states shown in Fig.~\ref{E} (right panel).
We use also the very well-tested localized
LSUB$m$ truncation scheme in which all multi-spin correlations 
are retained in the CCM correlation operators over all distinct locales on
the lattice defined by $m$ or fewer contiguous sites. The method of
solving for higher orders of LSUB$m$ approximations is discussed in
detail in Refs.~\cite{ccm2,ccm3}.
The number of independent spin configurations taken into account in the correlation operator 
increases rapidly with the truncation index
$m$. For example, at the highest
level of approximation used here, namely LSUB12, we take into account  750490 such
configurations.
 To obtain results in the exact $m \rightarrow \infty$ limit the raw
LSUB$m$ data must be extrapolated. Although there are no 
exact extrapolation rules, a great deal of experience has been garnered
by now for the gs energy and for the magnetic order parameter (sublattice
magnetization) $M$. 
For the gs energy per spin a well-tested and very accurate
extrapolation ansatz (see, e.g., Refs.,
(\cite{rachid05,schmalfuss,Bi:2008_PRB,darradi08,richter10,ccm3,Bi:2008_JPCM,reuther}) is
$E(m)/N = a_{0}+a_{1}m^{-2}+a_{2}m^{-4}\,,$  
while for the magnetic order parameter $M$ different schemes have been
employed for different situations \cite{extra}. 
Here we use 
$M(m) = b_{0}+b_{1}m^{-1/2}+b_{2}m^{-3/2}$. 
This scheme has been found to be appropriate for systems 
showing a gs order-disorder QPT (see, e.g.,
Ref.~\cite{Bi:2008_PRB,darradi08,richter10,Reuther:2011_J1J2J3mod}).
Since the hexagon is an important structural element of the honeycomb
lattice we decided to take into account for the extrapolations 
only LSUB$m$ data with $m \geq 6$.


CCM results for the gs energy per spin $E/N$ are shown in Fig.~\ref{E}
using both the N\'{e}el and collinear striped model states. They are
clearly well-converged for all values of $J_{2}$ shown. The
corresponding extrapolated LSUB$\infty$ results are also shown in the
regimes where real solutions exist for the entire data set, for each
choice of reference state. These results compare well to those from ED
calculations for $N=32$, also shown in Fig.~\ref{E}. The CCM results
show a clear intermediate region around $\kappa =0.5$ in which neither
of the quasiclassical AFM states is stable.  The ED results demonstrate a
``kink-like'' behavior in $E/N$ at $\kappa \approx 0.6$,
(cf. Refs.~\cite{Richter:2010_ED,Cabra}), which is a first hint of a
first-order QPT (and see the discussion below).

\begin{figure}
\includegraphics[width=5.2cm,angle=270]{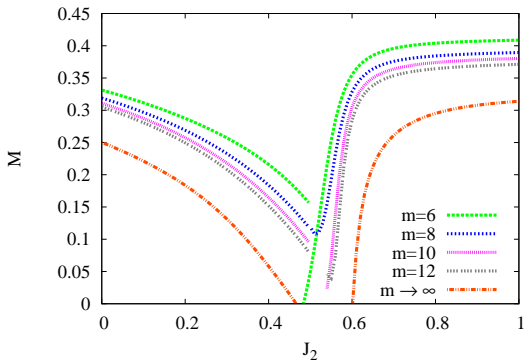}
\caption{
(Color online) CCM results for the gs order parameter, $M$ ($J_{1}
  \equiv 1$ and $J_{3}=J_{2}$) for 
the N\'{e}el and striped phases using LSUB$m$ ($m=6,8,10,12$) and extrapolated $m \rightarrow \infty$ (see text). 
}
\label{M}
\end{figure}
\begin{figure}
\includegraphics[width=5.2cm,angle=270]{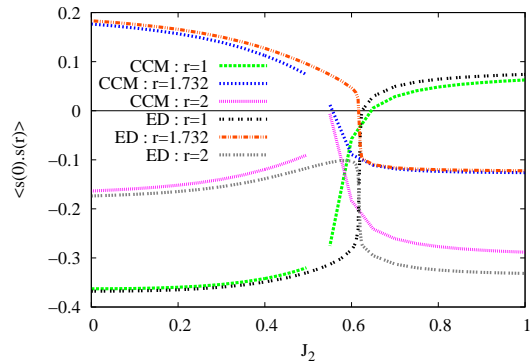}
\caption{(Color online) CCM LSUB12 and $N=32$ ED results for the
  spin-spin correlation function $\langle{\bf s}(0)\cdot{\bf
    s}(r)\rangle$ ($J_{1}
  \equiv 1$ and $J_{3}=J_{2}$).
Values $r=1,1.732,2$
  correspond respectively to nn, nnn, and nnnn pairs of spins.
 }
\label{CorrFunctn}
\end{figure}

We now discuss the magnetic order parameter $M$ in order to
investigate the stability of quasiclassical magnetic LRO. CCM results
for $M$ are shown in Fig.~\ref{M}.  The extrapolated N\'eel order
parameter vanishes at $\kappa_{c_{1}} \approx 0.466$, whereas the
corresponding collinear striped order parameter goes to zero at
$\kappa_{c_{2}} \approx 0.601$. These values can be considered as CCM
estimates for the QCPs of the model~\cite{comment}.  They are in good
agreement with those recently obtained by the Schwinger-boson approach
\cite{Cabra}, the functional renormalization group approach
\cite{honey7}, and by an ED approach \cite{lauchli2011}.  From the
shape of the order-parameter curve in Fig.~\ref{M} it seems probable
that the transition at $\kappa_{c_{1}}$ is continuous, whereas the
very steep fall near $\kappa_{c_{2}}$ may indicate a first-order
transition.  This observation is supported by the spin-spin
correlations functions shown in Fig.~\ref{CorrFunctn}, which also show
a similar ``smooth'' versus ``steep'' behavior near the points
$\kappa_{c_{1}}$ and $\kappa_{c_{2}}$, respectively.  Moreover, the
CCM and ED data presented in Fig.~\ref{CorrFunctn} are in good
quantitative agreement, which strongly supports the conclusions that
have been drawn above regarding the likely nature of the QPTs.

\begin{figure}[t]
\begin{center}
\mbox{
\subfloat{\includegraphics[width=6cm,height=6cm,angle=270]{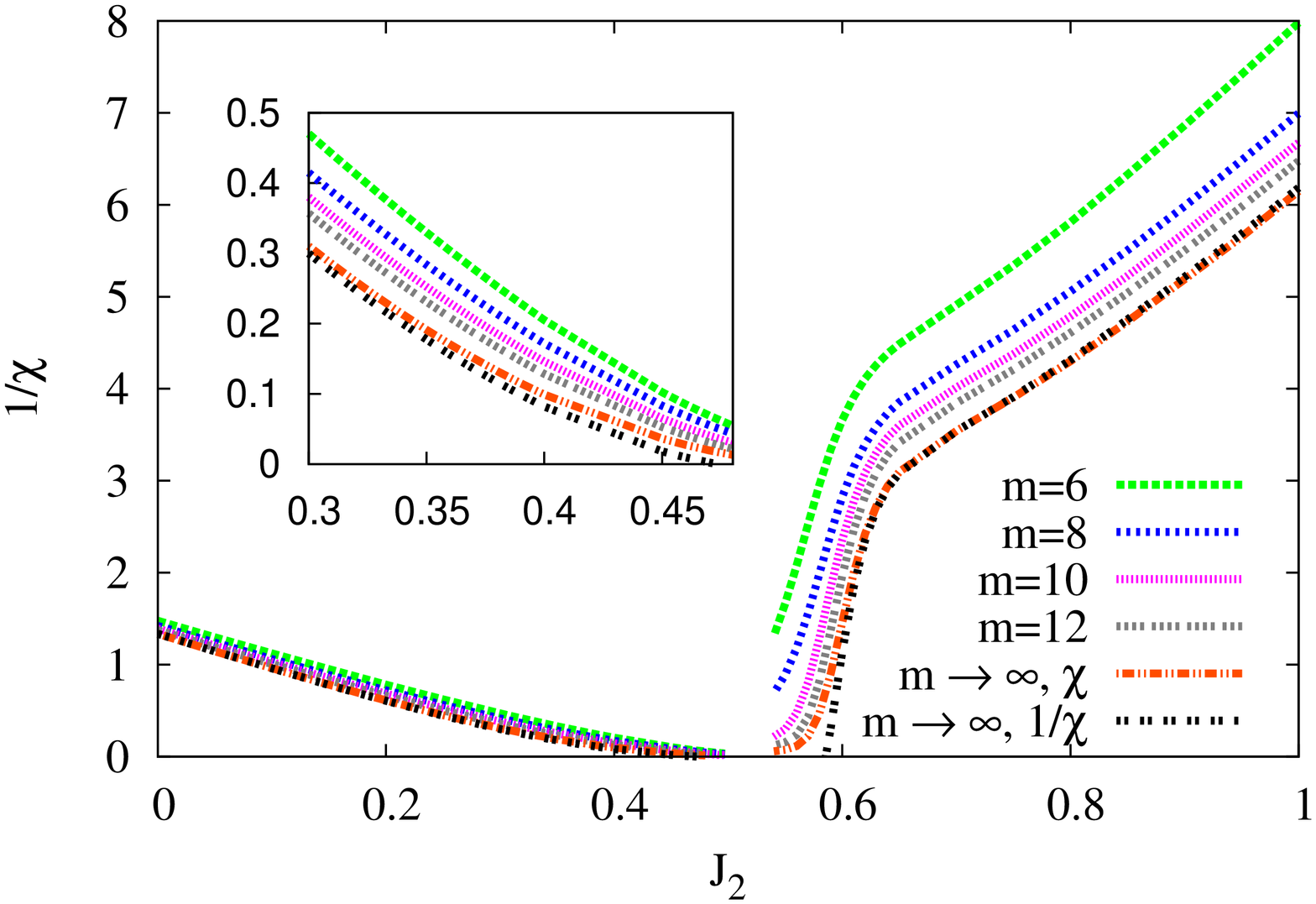}}
\raisebox{-3.5cm}{
\subfloat{\includegraphics[width=2.2cm,height=2.2cm]{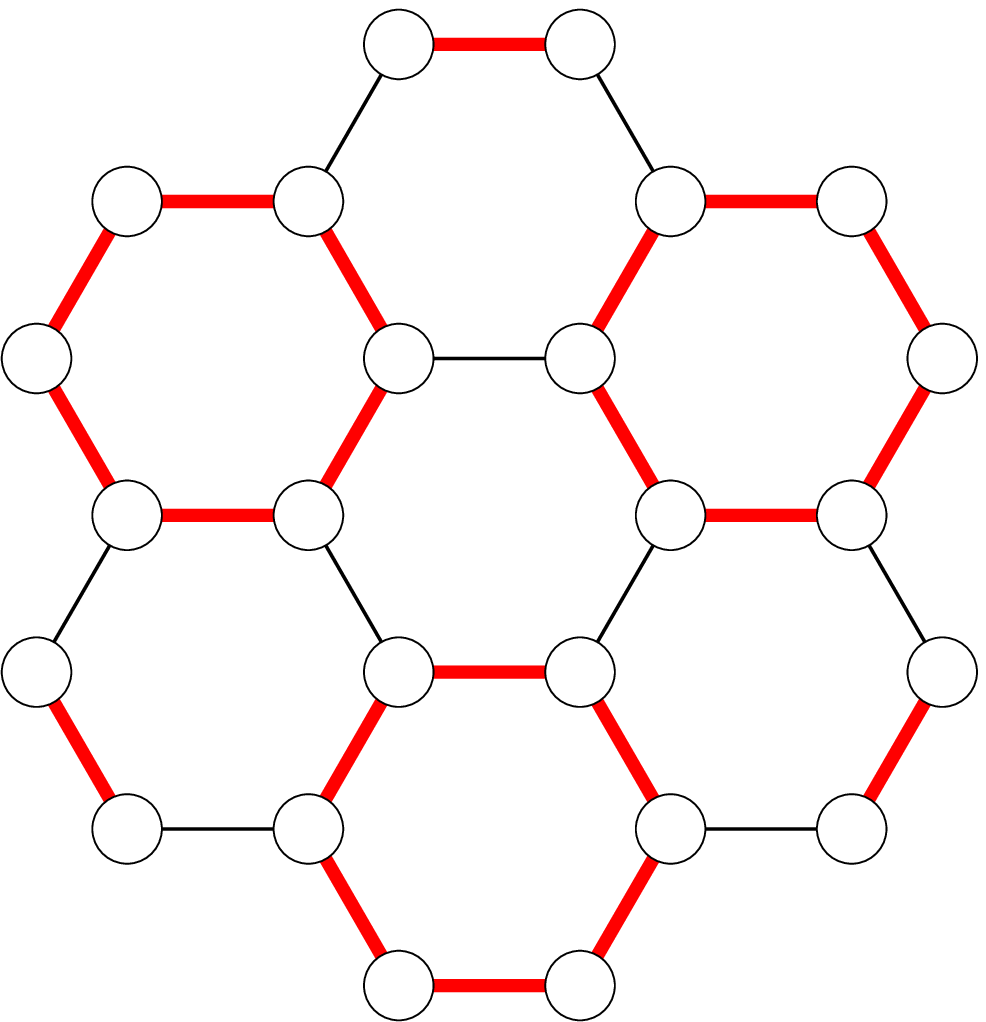}}
}
}
\caption{(Color online) Left: 
Results for the inverse plaquette susceptibility, $1/\chi$ ($J_{1} \equiv 1$ and $J_{3}=J_{2}$) via CCM LSUB$m$ ($m=6,8,10,12$), and extrapolated ($m \rightarrow \infty$) for both $\chi$ and $1/\chi$ (see text). 
Right: The perturbations (fields) $F=\delta\; \hat{O}$ for the plaquette susceptibility $\chi$.  
Thick red and thin black lines correspond respectively to strengthened and weakened
nn exchange couplings, where
$\hat{O} = \sum_{\langle i,j \rangle} a_{ij}
\mathbf{s}_{i}\cdot\mathbf{s}_{j}$, and the sum runs over all nn bonds, with
$a_{ij}=+1$ ($-1$) for thick red (thin black) lines.}
\label{X}
\end{center}
\end{figure}

Hitherto we have gathered strong evidence about the region of the
intermediate nonmagnetic quantum phase by using the high-order CCM and
ED techniques.  However, the question remains as to what is the nature
of this phase. We note that numerical evidence was found very recently
of a plaquette valence-bond crystal (PVBC) phase near $\kappa = 0.5$
\cite{lauchli2011}.  Note that such a PVBC phase is a strong candidate
for the QP phase of the $J_1$--$J_2$ model on the square lattice
\cite{zhito96,Sir:2006, darradi08}.  In order to analyze the
possibility of such a PBVC phase we first consider a generalized
susceptibility $\chi_F$ that describes the response of the system to a
``field'' operator $F$ (see, e.g., Refs.~\cite{Sir:2006,darradi08}).
We thus add a field term $F=\delta\; \hat{O}$ to the Hamiltonian
(\ref{eq1}), where $\hat{O}$ is an operator that in our case
corresponds to the possible PVBC order illustrated in Fig.~\ref{X}
(right panel), and which hence breaks the translational symmetry of
$H$.  We now calculate the energy per site $E(\delta)/N=e(\delta)$ for
the perturbed Hamiltonian $H+F$, by using the CCM for both the N\'eel
and the collinear striped reference states, and define the
susceptibility as $\chi_{F} \equiv -
\left. (\partial^2{e(\delta)})/(\partial {\delta}^2)
\right|_{\delta=0}$.  Rather less empirical experience is available
regarding the extrapolation of the CCM LSUB$m$ data for $\chi_{F}$
than for other quantities such as the gs energy or $M$. However, we
have tested several extrapolation schemes and found that the
extrapolation used for the gs energy, i.e. $\chi_{F}(m) =
c_{0}+c_{1}m^{-2}+c_{2}m^{-4}$, 
fits the data most accurately. 
For a crosscheck of this extrapolation scheme we also use corresponding extrapolation of the
inverse susceptibility, i.e. $\chi_{F}^{-1}(m) = d_{0}+d_{1}m^{-2}+d_{2}m^{-4}$,
and find consistent results as shown in Fig.~\ref{X}.
An instability of the gs against the perturbation $F$ is indicated by a divergence of $\chi_F$ or 
equivalently a zero point of $\chi_F^{-1}$. This fact might favor slightly 
the extrapolation of $\chi_F^{-1}$ when searching for such instabilities.  Our results
are in accordance with those of Ref.~\cite{lauchli2011} and they 
clearly favor a PVBC intermediate quantum phase instead of a structureless
spin-liquid phase. The extrapolated inverse susceptibility vanishes
on the N\'eel side at $\kappa \approx 0.473$ and on the collinear striped
side at $\kappa \approx 0.586$.
These values are in very good agreement with the critical values  $\kappa_{c_{1}} \approx 0.466$
and $\kappa_{c_{2}} \approx 0.601$ already found by the CCM order parameter. 
Thus, it is most likely that the PVBC phase occurs at (or is very close to) the 
transition points where the quasiclassical magnetic LRO breaks down.
Although there is a steep downturn of $\chi_F^{-1}$ at $\kappa_{c_{2}}$, we
find a smooth behavior for $\chi_F^{-1}$ at $\kappa_{c_{1}}$. This
evidence strongly supports our conclusion drawn from Figs.~\ref{E}, \ref{M}, 
and \ref{CorrFunctn} that the phase transition is of first-order type at $\kappa_{c_{2}}$ 
while that at $\kappa_{c_{1}}$ is of continuous type.

The possibility of deconfined quantum criticality for the frustrated
honeycomb HAFM was first pointed out in
Ref.~\cite{Sent:2004}. The first evidence supporting such a
scenario for the $J_1$-$J_2$-$J_3$ model on the honeycomb lattice has
been found very recently in Ref.~\cite{lauchli2011}.  Our CCM results
are seen to be highly converged and to agree well with those results
of the ED method.  Our results for the gs energy, magnetic order
parameter, spin-spin correlation function, and plaquette
susceptibility all point to collinear phases separated by a
magnetically disordered phase for $ 0.47 \lesssim \kappa \lesssim 0.60
$.  We have shown that the nature of this intermediate phase is
likely to be of plaquette valence-bond crystalline (PVBC)
type. Although we find indications of a first-order transition at
$\kappa_{c_{2}} \approx 0.60$, the transition between the N\'{e}el and
the PVBC phases at $\kappa_{c_{1}}$ appears continuous.  Since the
N\'eel and the PVBC phases break different symmetries and we have
shown that they are likely to meet at $\kappa_{c_{1}} \approx 0.47$,
our results present strong evidence that the frustrated
spin-$\frac{1}{2}$ HAFM on the honeycomb lattice contains a continuous
QPT described by the scenario of deconfined quantum criticality.

We thank the University of Minnesota Supercomputing Institute for the grant of
supercomputing facilities. 
J.R. thanks A. L\"auchli for fruitful
discussions.


\begin{thebibliography}{200}
\bibitem{extra}
Note that for unfrustrated models the  extrapolation 
$M(m) = b_{0}+b_{1}m^{-1}+b_{2}m^{-2}\,$ is more
appropriate \cite{ccm3}. 
This scheme gives the 
best result, $M
\approx 0.273$, for the
pure nn HAFM on the honeycomb lattice ($J_{2}=J_{3}=0, J_{1}=1$).
\bibitem{comment}
To estimate the accuracy within the used extrapolation scheme we have also
determined $\kappa_{c_{1}}$ and  $\kappa_{c_{2}}$ using LSUB$m$ with
$m=\{6,8,10\}$, which gives $\kappa_{c_{1}} \approx 0.448$
and $\kappa_{c_{2}} \approx 0.601$. The values
$\kappa_{c_{1}}$ and  $\kappa_{c_{2}}$ of course depend also on the extrapolation
scheme employed.
\end{thebibliography}
\end{document}